\newif \ifLong
\Longtrue  

\documentclass[conference,letterpaper,10pt]{IEEEtran}

\newif \ifisit
\isitfalse

\usepackage[table]{xcolor}
\usepackage{times}
\usepackage{epsfig}
\usepackage[cmex10]{amsmath}
\usepackage{amsthm}
\usepackage{amsfonts}
\usepackage{graphicx}
\usepackage{amssymb}
\usepackage{amstext}
\usepackage{latexsym}
\usepackage{color,colortbl}
\usepackage{ifthen}
\usepackage{multirow}
\usepackage{verbatim}
\usepackage{array,tabularx}
\usepackage{arydshln}
\usepackage[mathscr]{euscript}
\usepackage{accents}
 \usepackage{cite}
\usepackage{hhline}
\usepackage{caption}
\usepackage{subcaption}
\usepackage{enumerate}
\usepackage{xcolor}
\usepackage{mathtools}
\usepackage{url}
\usepackage{xparse}
\usepackage{makecell}
\usepackage{varwidth}
\usepackage{bm}
\usepackage{arydshln}

\usepackage{hyperref}
\usepackage[shortlabels]{enumitem}
\usepackage{cleveref}
\hypersetup{
    colorlinks=true,
    linkcolor=blue,
    filecolor=magenta,
    urlcolor=cyan,
    pdftitle={Overleaf Example},
    pdfpagemode=FullScreen,
}

\usepackage{comment}
\usepackage{wrapfig}

\usepackage{caption}
\usepackage{float}
\usepackage{booktabs}

\usepackage{dirtytalk}

\usepackage{mathtools}

\usepackage{booktabs}

\usepackage{float}

\usepackage{multirow}

\newcolumntype{C}[1]{>{\centering\let\newline\\\arraybackslash\hspace{0pt}}m{#1}}

\newcommand{\cry}{\operatorname{Crypt}}
\newcommand{\pad}{\operatorname{Pad}}

\usepackage[normalem]{ulem}

\usepackage{amsmath,pgfplots,amssymb}
\usepackage{graphicx}

\newtheorem{theorem}{Theorem}

\newtheorem{proposition}{Proposition}

\theoremstyle{definition}

\newtheorem{definition}{Definition}

\theoremstyle{definition}

\theoremstyle{definition}

\newcommand{\interior}[1]{%
  {\kern0pt#1}^{\mathrm{o}}%
}

\usetikzlibrary{matrix,decorations.pathreplacing}

\definecolor{DarkGreen}{rgb}{0.1,0.5,0.1}
\definecolor{DarkRed}{rgb}{0.5,0.1,0.1}
\definecolor{DarkBlue}{rgb}{0.1,0.1,0.5}
\definecolor{DarkPurple}{rgb}{0.5,0.2,0.5}
\definecolor{DarkTurquoise}{rgb}{0.1,0.5,0.5}

\definecolor{beaublue}{rgb}{0.74, 0.83, 0.9}
\definecolor{coolblack}{rgb}{0.0, 0.18, 0.39}
\definecolor{apricot}{rgb}{0.98, 0.81, 0.69}
\definecolor{burntorange}{rgb}{0.8, 0.33, 0.0}
\definecolor{blue-violet}{rgb}{0.54, 0.17, 0.89}
\definecolor{byzantium}{rgb}{0.44, 0.16, 0.39}
\definecolor{brilliantrose}{rgb}{1.0, 0.33, 0.64}
\definecolor{cerisepink}{rgb}{0.93, 0.23, 0.51}
\definecolor{cobalt}{rgb}{0.0, 0.28, 0.67}
\definecolor{bostonuniversityred}{rgb}{0.8, 0.0, 0.0}

\definecolor{ao(english)}{rgb}{0.0, 0.5, 0.0}

\usepackage{nidanfloat}

\usepackage[utf8]{inputenc}
\usepackage[T1]{fontenc}
\usepackage{url}
\usepackage{ifthen}
\usepackage{cite}

\newcommand{\off}[1]{}

\begin{document}



\title{AES as Error Correction:\\ Cryptosystems for Reliable Communication}


\author{Alejandro Cohen\IEEEauthorrefmark{1}, Rafael G. L. D'Oliveira\IEEEauthorrefmark{2}, Ken R. Duffy\IEEEauthorrefmark{3}, Jongchan Woo\IEEEauthorrefmark{4}, Muriel Médard\IEEEauthorrefmark{4}   \\
\IEEEauthorrefmark{1}Faculty of Electrical and Computer Engineering, Technion, Israel, Email: alecohen@technion.ac.il\\
\IEEEauthorrefmark{2}SMSS, Clemson University, USA, Email: rdolive@clemson.edu\\
\IEEEauthorrefmark{3}RLE, Massachusetts Institute of Technology, USA, Emails:  \{jc\_woo, medard\}@mit.edu\\
\IEEEauthorrefmark{4}Hamilton Institute, Maynooth University, Ireland, Email:  ken.duffy@mu.ie \off{\vspace{-0.5cm}}}


\maketitle

\begin{abstract}
\ifLong
In this paper, we show that the Advanced Encryption Standard (AES) cryptosystem can be used as an error-correcting code to obtain reliability over noisy communication and data systems. Moreover, we characterize a family of computational cryptosystems that can potentially be used as well performing error correcting codes. In particular, we show that simple  padding followed by a cryptosystem with uniform or pseudo-uniform outputs can approach the error-correcting performance of random codes. We empirically contrast the performance of the proposed approach using AES as error correction with that of Random Linear Codes and CA-Polar codes and show that in practical scenarios, they achieve almost the same performance. Finally, we present a modified counter mode of operation, named input plaintext counter mode, in order to utilize AES for multiple blocks while retaining its error correcting capabilities.
\fi
\end{abstract}




\section{Introduction}
Since Shannon's revolutionary work in 1948 \cite{shannon1948mathematical} to modern data systems \cite{lin20215g}, encoding with error correcting codes is used to obtain reliable communication at the highest possible data rates
\cite{huffman2021concise,cover2012elements}. As for security, common computational cryptosystems, like the Advanced Encryption Standard (AES) \cite{dworkin2001advanced,nechvatal2001report}, are used in a pre-processing stage in order to encrypt the information \cite{hoffstein2008introduction,stallings2006cryptography}. Thus, the traditional encoding process consists of two separated stages/systems: encryption followed by an error correcting code.  More recently, however, the reverse ordering of encoding and then encrypting was proposed in a setting where reliability and security can be obtained by partial encryption after error correction encoding \cite{cohen2022partial}.

\begin{figure}[!t]
    \centering
    \ifLong
    \includegraphics[trim={7.2cm 5.8cm 7cm 8cm},clip, width = 1\columnwidth]{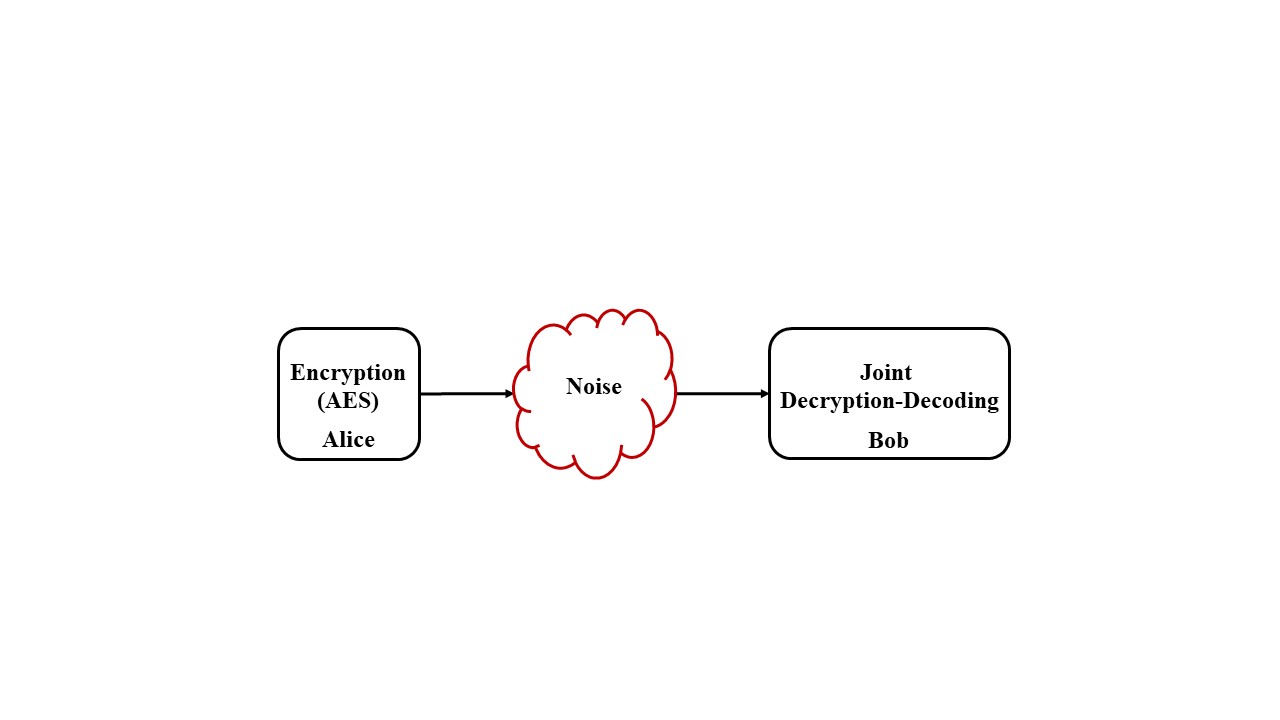}
    \fi
    \caption{Reliable communication over noisy channel, one source, Alice, one legitimate destination, Bob.}
    \label{fig:Sys}
\end{figure}

The interplay between coding and cryptography has a long and deep history. Indeed, uniformly distributed outputs appear both in the coding scheme of the original work of Shannon and as a desideratum for cryptosystems, albeit for different purposes. In this work, we characterize a family of computational cryptosystems that can be potentially as error correcting codes. We show that, by utilizing a mere padding in order to manage the code rate, followed by an encryption with a cryptosystem with uniformly distributed outputs, we can build a high performance error correction mechanism. Thus, as illustrated in Figure~\ref{fig:Sys}, the encoding can be performed in a single stage by the cryptosystem while achieving reliability at high data rates. Cryptosystems that have outputs that are random or pseudo-random, say AES or as in \cite{moller2004public}, can serve as error correction codes that can approach capacity.

We empirically demonstrate our idea by showing that a simple padding followed by AES, is an effective mechanism for error correction when combined with Guessing Additive Noise Decoding (GRAND) \cite{duffy2019capacity, duffy19GRAND, duffy2021ordered}, which can decode any code, including the nonlinear construct yielded by AES. Our joint decryption-decoding scheme is based on a recently proposed decoding scheme \cite{cohen2022partial}. Moreover, the performance of our error-correction scheme, utilizing padding and AES, is similar to that of Random Linear Codes (RLC), which are known to be capacity-achieving \cite{gallager1973random}, and with CRC-Aided Polar (CA-Polar) codes \cite{tal2015list}.

\ifLong
In application, AES is combined with a block cipher mode \cite{ferguson2011cryptography}. Common modes, like cipher block chaining (CBC), suffer from error propagation, while counter (CTR) mode, because it does not input the plaintext into AES, is not well suited for our technique. To circumvent these issues we present a modified version of CTR, named input plaintext counter (IP-CTR) mode. As opposed to the standard CTR mode, in IP-CTR mode the plaintext is added to the counter before the AES encryption.

The paper is organized as follows. In Section~\ref{sec:system}, we provide the setting and problem formulation, and we provide some definitions and metrics. In Section~\ref{sec:cry_ec}, we show how to utilize cryptosystems with
uniform output for error correction. In Section~\ref{sec:main}, we present the AES as an error correction code scheme and empirically demonstrate its performance. In Section~\ref{sec:modes}, we present a new block cipher mode in order to utilize AES for error correction. Finally, we summarize the paper in Section~\ref{sec:conc} and discuss possible future directions.
\else\fi

\section{Preliminaries}\label{sec:system}

\subsection{Setting}

We consider a setting where a transmitter, Alice, wishes to transmit a message $M \in \mathbb{F}_{2}^{k}$ to a legitimate receiver, Bob, over a noisy channel. For the noisy communication at Bob, we consider two models. An independent binary symmetric channel (BSC) with a bit flip probability of $p<\frac{1}{2}$, and an Additive white Gaussian noise (AWGN) channel with noise~$N$. The noise $N\sim\mathcal{N}\left(0,\sigma^2\right)$ is independent and identically distributed and drawn from a zero-mean normal distribution with variance $\sigma^2$ \cite{cover2012elements}. We denote by $X$ and $Y$ the encoded message that Alice transmits to Bob and the noisy observation at Bob, respectively.

\begin{figure*}[!t]
    \centering
    \includegraphics[trim={0cm 3.5cm 0cm 1.5cm},clip, width = 1\textwidth]{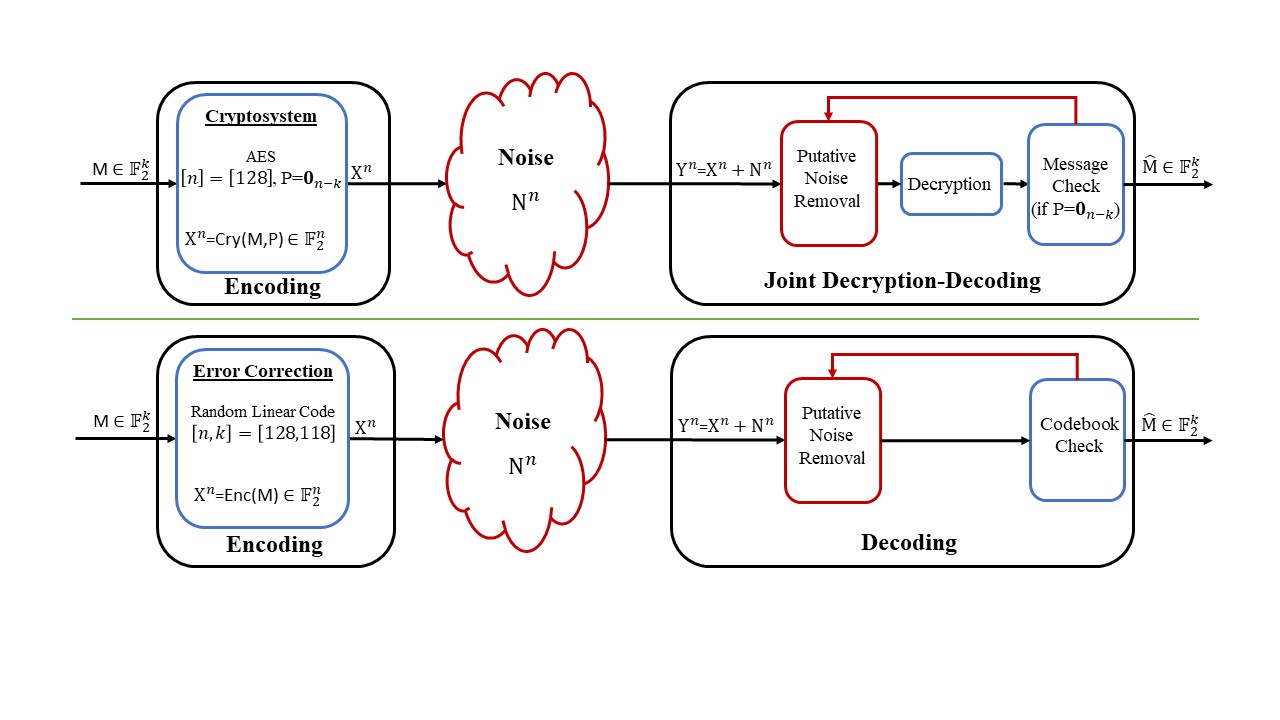}
    \caption{Proposed cryptosystem with a uniform outcome as error correction, e.g., AES (upper figure) and an RLC decoded with GRAND (bottom figure).}
    \label{fig:scheme}
\end{figure*}

\subsection{Cryptosystems}

A cryptographic system is a set of algorithms which converts messages (plaintexts) into cyphertexts in order to transmit them securely through a communication channel. We consider both asymmetric and symmetric cryptosystems.

\begin{definition}[Cryptosystem]\label{Crypto_scheme}
A cryptosystem consists of three algorithms:

\begin{itemize}
        \item A key generation algorithm $\mathrm{Gen}(\kappa)$ which takes as input a security parameter $\kappa$ and generates a public key $p_k$ and a secret key $s_k$, in the case of an asymmetric cryptosystem. In the case of a symmetric cryptosystem the key generation algorithm outputs a single key. For simplicity of notation we denote this single key by $p_k=s_k$.
        \item An encryption algorithm $\cry_{p_k}(m)$ which takes as input a message $m$ belonging to some set of messages $\mathcal{M}$ and the public key $p_k$ and then outputs a ciphertext $c$ belonging to some set of ciphertexts.
        \item A polynomial time decryption algorithm $\mathrm{Dec}_{s_k}(c)$ which takes as input a ciphertext $c=\cry_{p_k}(m)$ and the secret key $s_k$ and outputs the original message $m$.
    \end{itemize}

\end{definition}

Since in our communication model we are working over bits, we restrict the inputs and outputs of our encryption algorithm to binary vectors, i.e., our symmetric cryptosystem encryption is of the form $\cry_{p_k} : \mathbb{F}_2^n \rightarrow \mathbb{F}_2^{n}$. Moreover, to allow for probabilistic cryptosystems, we consider encryption algorithms of the form $\cry_{p_k} : \mathbb{F}_2^n \times \mathbb{F}_2^{r_0} \rightarrow \mathbb{F}_2^{n+r}$, with $r_0 \leq r$, where $r_0$ corresponds to the amount of random bits ($r_0=0$ for deterministic schemes) used by the algorithm and $r$ corresponds to a possible increase in the output space. Thus, if Alice wishes to send a message $M \in \mathbb{F}_2^n$, she chooses $\bar{R} \in \mathbb{F}_2^{r_0}$ uniformly at random and then transmits $\cry_{p_k} (M,\bar{R}) \in \mathbb{F}_2^{n+r}$ to Bob.

\subsection{Random Functions and Random Codes}
The connection between random functions and random codes, together with the noisy-channel coding theorem \cite{shannon1948mathematical} considered in this subsection, is the key to our main result given in Section~\ref{sec:cry_ec}.

\begin{definition}[Random Function] \label{def: random function}
A random function is a function $f: \mathbb{F}_2^k \rightarrow \mathbb{F}_2^n$ chosen uniformly at random from the set of all functions from $\mathbb{F}_2^k$ to $\mathbb{F}_2^n$.
\end{definition}
It is important to note that a pseudorandom function, as usually utilized in classical cryptosystems, is computationally indistinguishable from a random function \cite{goldwasser2008lecture}.

\begin{definition}[Random Code] \label{def: random code}
Let $k,n \in \mathbb{Z}_+$ be positive integers. A random code with rate $k/n$ is a subset $\mathcal{C} \subseteq \mathbb{F}_2^n$ chosen uniformly at random from the set of all subsets of $\mathbb{F}_2^n$ with cardinality less or equal than $2^k$.
\end{definition}

In the asymptotic regime, i.e., when $n$ goes to infinity, random codes are capacity achieving for the BSC and AWGN channels. In particular, this implies that as $n$ goes to infinity, the error probability of a code chosen uniformly at random with fixed rate $R=\frac{k}{n}<C$ goes to zero, where $C$ is the capacity of the underlying BSC or AWGN channel.

\begin{proposition}[Random Codes are Good] \label{prop: random codes good}
Let $C$ be the capacity of the underlying BSC or AWGN channel, and $\mathcal{C}$ be a random code with rate $R=\frac{k}{n}<C$. Then, as $n\rightarrow \infty$, the error probability $P_e(\mathcal{C})\rightarrow 0$.
\end{proposition}

\begin{proof}
The noisy-channel coding theorem \cite{shannon1948mathematical} states that in the asymptotic regime when $n \rightarrow \infty$, random codes with rate lower than the channel capacity can reliably transmit with an arbitrarily low probability of error,
\begin{equation}\label{eq:err_pe}
    P_e \leq \varepsilon + 2^{3n\varepsilon}2^{-n(C-R)},
\end{equation}
for any $R < C -3\varepsilon$ \cite[Chapter 7.7]{cover2012elements}. Thus, for $R<C$, one can choose $\varepsilon$ and $n$ such that the average probability of error averaged over random codes is less than $2\varepsilon$.

The deviation from the probability of error, averaged over random codes, is shown to be small in the literature \cite{1056514,tamir2020large}. For our needs, this follows from Markov's inequality, and shows that bad random codes are rare. For a random code $\mathcal{C}$ with rate $R=\frac{k}{n}<C$ and any $c>1$, the deviation on the probability of error is bounded by
\[
 P(P_e(\mathcal{C})\geq 2c\varepsilon) \leq \frac{\mathbb{E}[P_e(\mathcal{C}]}{2c\varepsilon}\leq \frac{2\varepsilon}{2c\varepsilon}=\frac{1}{c},
\]
where the second inequality is given by the noise channel coding theory, such that $\mathbb{E}[P_e(\mathcal{C})]\leq 2\varepsilon$. Thus, to obtain a low deviation, $c$ must be large, while $2c\varepsilon$ is small. Since $\varepsilon$ can be made arbitrarily small for large enough $n$, as shown in \eqref{eq:err_pe}, it follows that for large $n$ the deviation can be made arbitrarily small. Thus, for large $n$, random codes are good with high probability.
\end{proof}

The key point we utilize in the next section is that if $f:\mathbb{F}_2^k \rightarrow \mathbb{F}_2^n$ is a random function, then $\mathcal{C} = f(\mathbb{F}_2^k) \subseteq \mathbb{F}_2^n$ is a random code with rate $R = k/n$. Thus, as shown in Proposition \ref{prop: random codes good}, the code $\mathcal{C} = f(\mathbb{F}_2^k)$ has a high probability of being a good code. In Section \ref{sec:main} we show how to decode utilizing a modified version of a joint decryption-decoding scheme with GRAND decoder \cite{duffy2019capacity,duffy2021ordered} presented in \cite[Section VI]{cohen2022partial}.

\off{\textcolor{blue}{\begin{definition}
Let $k,n \in \mathbb{Z}_+$ be positive integers. A random code with rate $k/n$ is a subset $\mathcal{C} \subseteq \mathbb{F}_2^n$ chosen uniformly at random from the set of all subsets of $\mathbb{F}_2^n$ with cardinality less or equal than $2^k$.
\end{definition}}}


\section{Cryptosystems as Error Correcting Codes}\label{sec:cry_ec}

In this section, we show how to utilize cryptosystems with uniform output for error correction. Recall that Alice wants to transmit a message $M \in \mathbb{F}_2^k$ to Bob through a noisy communication channel. Our Cryptosystem as Error Correcting (CEC) scheme consists in first injectively embedding the message into a larger space by means of an injective function $\pad : \mathbb{F}_2^k \rightarrow \mathbb{F}_2^n.$
Any injective function works, so for simplicity, we can think of $\pad$ as the function which pads the message with $n-k$ zeros\footnote{In terms of security guarantees, further analysis is warranted. We leave this interesting direction for future work.}.

Suppose we are given a cryptosystem with an encryption algorithm of the form $\cry_{p_k} : \mathbb{F}_2^n \times \mathbb{F}_2^{r_0} \rightarrow \mathbb{F}_2^{n+r}$,
satisfying Definition~\ref{def: random function}. Then, in the case of deterministic cryptosystems, i.e., when $r_0=r=0$, we have the following result, which follows using CEC scheme from combining Definitions \ref{def: random function} and \ref{def: random code}, and Proposition \ref{prop: random codes good}.\off{\textcolor{blue}{The noisy-channel coding theorem states that in the asymptotic regime when $n \rightarrow \infty$, random codes with rate lower than the channel capacity can reliably transmit with an arbitrarily low probability of error.}}

\off{\begin{theorem}\label{tho:deterministic}
Let $C_{p}$ be the capacity of the binary symmetric channel with crossover probability $p$, and suppose that the output of $\cry$ is uniformly distributed. Then, if $\frac{k}{n}<C_{p}$, deterministic CEC scheme can asymptotically transmit with an arbitrarily low probability of error at a rate of $\frac{k}{n}$.
\end{theorem}}

\begin{theorem}\label{tho:deterministic}
Let $C$ be the capacity of the underlying BSC or AWGN channel, and suppose that $\cry_{p_k}$ is a random function. Then, if $\frac{k}{n}<C$, the deterministic CEC scheme can asymptotically transmit with an arbitrarily low probability of error at a rate of $\frac{k}{n}$.
\end{theorem}

\begin{proof}
Alice is transmitting codewords from the code
\begin{align*}
    \mathcal{C} = \cry_{p_k} \left( \pad (\mathbb{F}_2^k) \right) \subseteq \mathbb{F}_2^{n}
\end{align*}
through the channel. This code has a rate of $\frac{k}{n}$. Since in the CEC scheme the cryptosystem $\cry$ \off{\textcolor{red}{can be only an injective function of the entire input, i.e., the message with the padding together, and}}is a random function as given in Definition~\ref{def: random function}, the code $\mathcal{C}$ is a random code as given in Definition~\ref{def: random code}. Note that in the CEC scheme, the code $\mathcal{C}$ is a random code used to decode correctly the message together with the padding bits. Thus, no information is gained from the padding bits as they are used solely as a redundancy check \cite{peterson1961cyclic}. It then follows from Proposition~\ref{prop: random codes good}, using the noisy-channel coding theorem, that if $\frac{k}{n}<C$, then the code $\mathcal{C}$ can asymptotically transmit with an arbitrarily low probability of error, i.e., $P_e (\mathcal{C})\rightarrow 0$ as $n \rightarrow \infty$.
\end{proof}

In the case of probabilistic cryptosystems $\cry_{p_k}$, i.e., when $0 < r_0\leq r$, we obtain the following result.

\begin{theorem}\label{coro:capacity}
Let $C$ be the capacity of the underlying BSC or AWGN channel, and suppose that $\cry_{p_k}$ is a random function. Then, if $\frac{k+r_0}{n+r}<C$, the probabilistic CEC scheme can asymptotically transmit with an arbitrarily low probability of error at a rate of $\frac{k}{n+r}$.
\end{theorem}

\begin{proof}
The proof follows directly by using the same arguments as in Theorem~\ref{tho:deterministic}. However, in this case, Alice is transmitting codewords from the code
\begin{align*}
    \mathcal{C} = \cry_{p_k} \left( \pad (\mathbb{F}_2^k) \times \mathbb{F}_2^{r_0} \right) \subseteq \mathbb{F}_2^{n+r}
\end{align*}
through the noisy-channel. In this probabilistic scheme, no information is gained from the $r_0$ random bits. Thus, the probabilistic CEC scheme has a communication rate of $\frac{k}{n+r}$.

In the probabilistic CEC scheme for every message $M \in \mathbb{F}_2^k$ the code $\mathcal{C}$ has $2^r$ codewords corresponding to $M$. Thus, when utilizing the noisy-channel coding theorem as given in Proposition~\ref{prop: random codes good}, we do not need to declare an error if a ball around the received transmission has codewords corresponding to the same message.
\end{proof}

Note, that the bound $\frac{k+r_0}{n+r}<C$ in Theorem~\ref{coro:capacity} might not be tight, i.e., higher rates than those stated in the theorem might be achievable. An interesting direction to investigate is what occurs if instead of utilizing a zero padding, we utilize $r_0$ random bits as the padding.

\section{AES as Error Correction}\label{sec:main}
In this section, we present the CEC scheme utilizing the Advanced Encryption Standard (AES) \cite{dworkin2001advanced,nechvatal2001report} as an error-correcting code. In Figure~\ref{fig:scheme}, we illustrate the encoding and decoding operations for the proposed scheme using AES as an error correcting code, and for a traditional error correction scheme using an RLC code \cite{gallager1973random}. The noise entropy we consider for the noisy communication at Bob is $H(p)$ for the BSC channel and $h(N)$ for the AWGN channel \cite{cover2012elements}. Let $\nu$ denote the entropy of the noise according to the particular channel considered. Thus, to obtain reliable communication we consider messages with size $k<n - \nu$. For the AES cryptosystem, we use a standard FIPS-197\off{ $n=128$-bit} scheme as given in \cite{pub2001197}. This is a common cryptosystem that belongs to the family of cryptosystems that can serve as error-correcting as defined in Section~\ref{sec:cry_ec}.

We start by presenting the encoding process at Alice. A message $m \in \mathbb{F}_2^k$ of size $k$ is encrypted by concatenating a padding $P$ of size $n-k$. For simplicity, the AES scheme considered does not include extra random symbols in the encryption process, i.e, $r_0=r=0$. Thus the rate in the proposed scheme is $R=\frac{k}{n}$ and encoder at Alice is given by
\begin{equation*}
    \text{AES Crypt}: \mathbb{F}_2^k \times \mathbb{F}_2^{n-k} \rightarrow \mathbb{F}_2^n.
\end{equation*}
Any padding $P$ of size $n-k$ serves our purpose. For simplicity, we set $P=\textbf{0}_{n-k}$, the all zero vector of length $n-k$.

At Bob, given the noisy observation $Y \in \mathbb{F}_2^n$, the joint decryption-decoder is of the form
\begin{equation*}
    \text{Joint-2-Dec}: \mathbb{F}_{2}^{n} \rightarrow \mathbb{F}_{2}^{k},
\end{equation*}
and outputs an estimate $\hat{M} = \text{Joint-2-Dec}(Y)$,
of the original message $M \in \mathbb{F}_2^k$. To decode and decrypt the message, Bob utilizes a modified version of a joint decryption-decoding scheme with GRAND decoder \cite{duffy2019capacity,duffy2021ordered} as presented in \cite[Section VI]{cohen2022partial}. Given the noisy observation Y, the decoder orders the noise sequences from the most likely to least likely. Then, the decoder goes through the list, subtracting the noise from $Y$ and performing an AES decryption. At the last stage, the decoder checks if the last $n-k$ padding bits of $P$ are correct, in the case considered here, if they are all zero. The first time all the padding bits are correct, Bob declares the first $k$ bits as the message transmitted. We refer the readers to \cite{cohen2022partial} for a detailed description and analysis of the joint decryption-decoding scheme with GRAND decoder.

\begin{figure*}[!t]
    \centering
    \includegraphics[width= 0.49\textwidth]{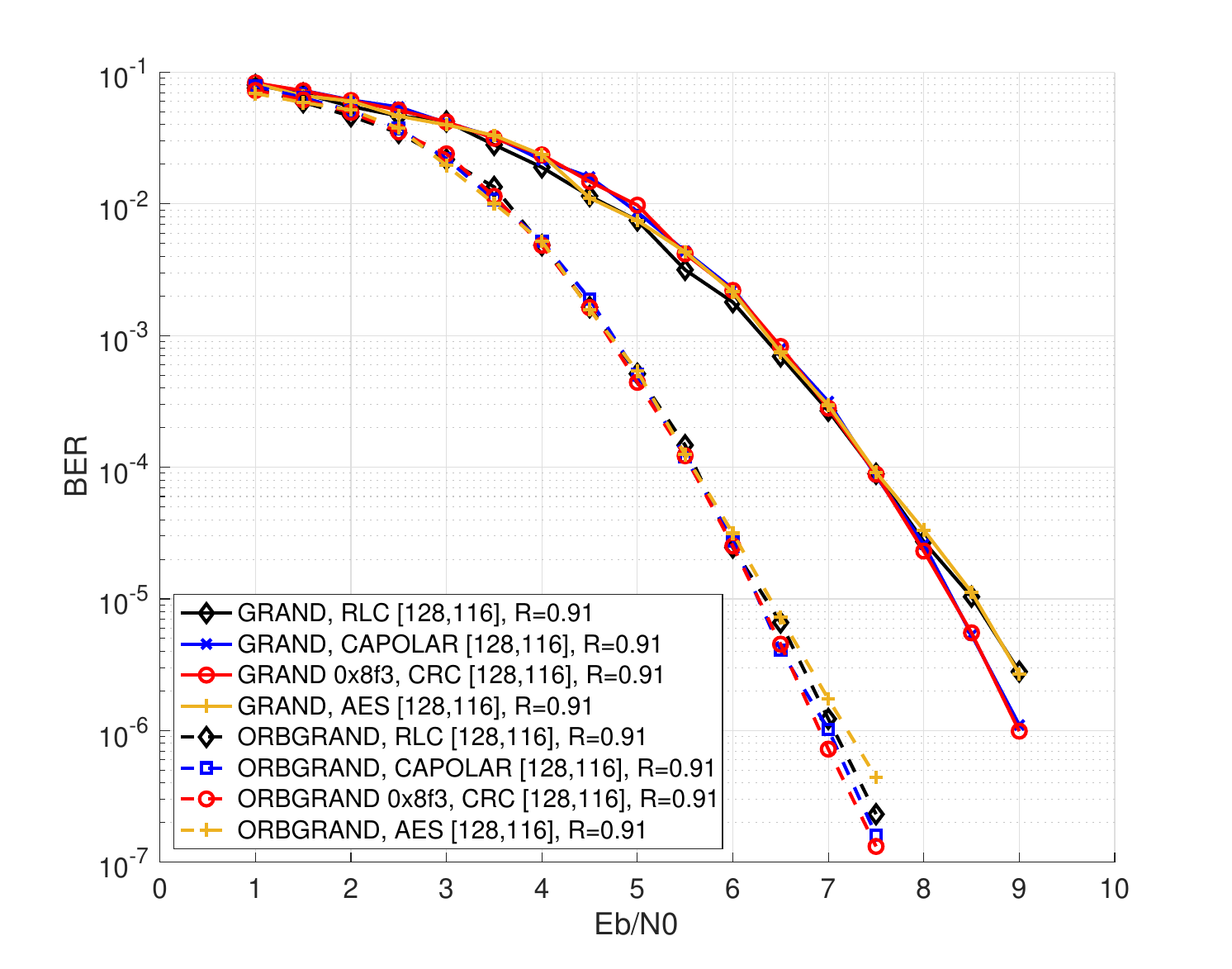}
    \includegraphics[width= 0.49\textwidth]{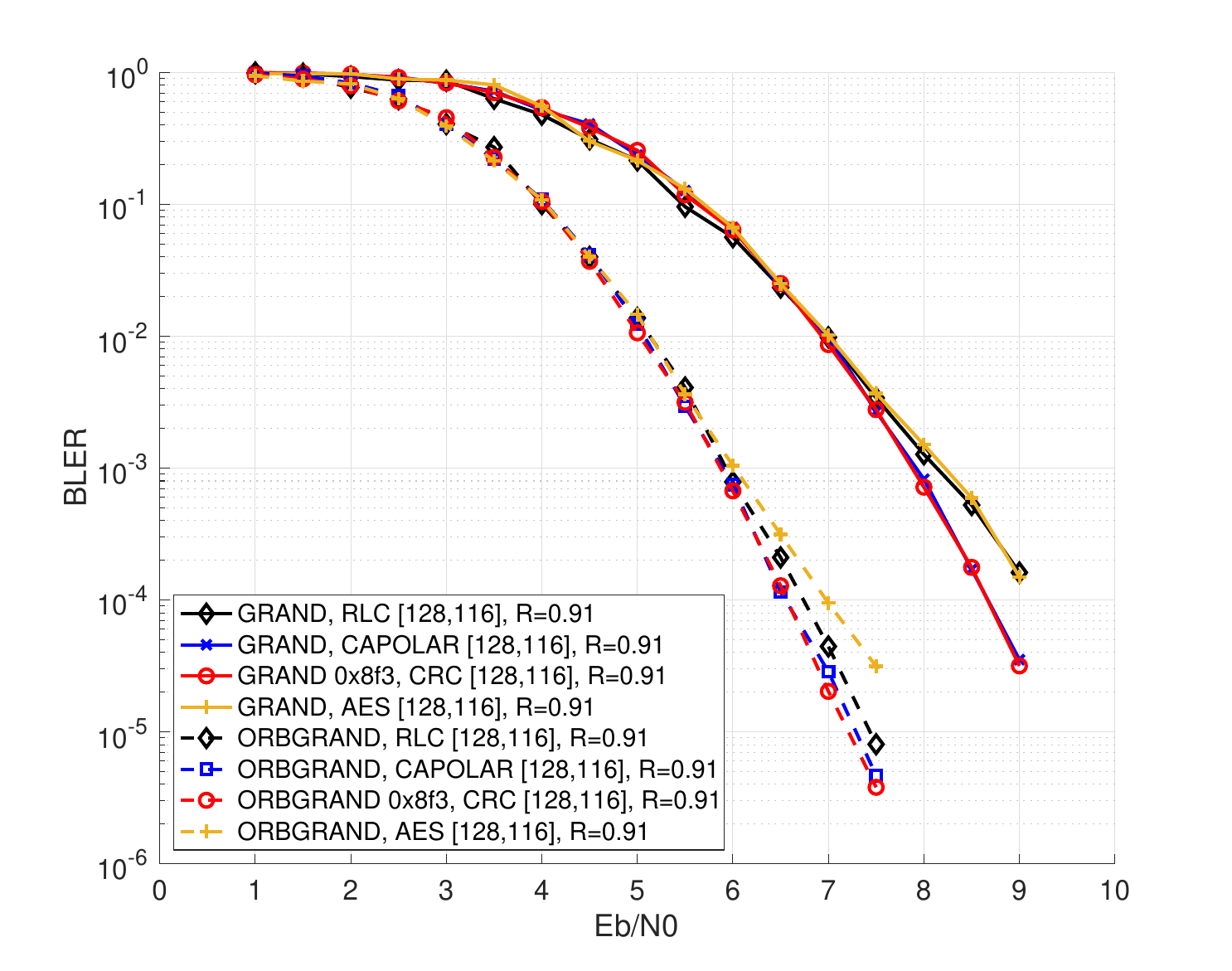}
    \caption{BER on the left and BLER on the right vs. Eb/N0 for codes of length n = 128 and rate R = 0.91 encoded with AES as error correcting code and with RLC and CA-Polar code. The joint decryption-decoding is performed with GRAND and ORBGRAND with soft-information.}
    \label{fig:sim}
\end{figure*}

In Subsection~\ref{subsec:sim}, we empirically demonstrate the performance at Bob with the scheme proposed and compare that with traditional solutions using RLC and CA-Polar code, as are illustrated in Figure~\ref{fig:sim}.

\subsection{Empirical study}\label{subsec:sim}
In this subsection, we demonstrate the performance of AES as an error-correcting code. We empirically validate the theoretical result given in Section~\ref{sec:cry_ec}. That is, we show that in practical scenarios, a cryptosystem such as AES, whose outcome is considered pseudo-random, has an error-correcting performance which is similar to random linear codes (RLCs). Indeed, we show that the bit error rate (BER) and block error rate (BLER) performance of AES as an error-correcting code is almost the same as those of an RLC and CA-Polar code. Our simulations employ a Gaussian channel model with AWGN and Binary Phase Shift Keying (BPSK) modulation. For encoding codes, we use:
\begin{itemize}
\item Traditional AES-FIPS-197 \cite{dworkin2001advanced,nechvatal2001report} with zero padding as proposed in Section~\ref{sec:main}.
\item
Random linear codes (RLCs), which have long since been known to be
capacity achieving~\cite{gallager1973random}, but have been little
investigated owing to the lack of availability of decoders until
the recent development of guessing random additive noise decoding
(GRAND) \cite{duffy2018guessing,duffy2019capacity}.
\item
CRC-aided polar (CA-Polar) Codes~\cite{tal2011list,tal2015list},
which are Polar codes~\cite{arikan2008channel,arikan2009channel}
with an outer cyclic redundancy check (CRC) code. These have been proposed for all control
channel communications in 5G NR~\cite{3gppcapolar1}.
\end{itemize}

For decoding, we use a modified version of a joint decryption-decoding scheme with the GRAND decoder \cite{duffy2018guessing,duffy2019capacity} as presented in Section~\ref{sec:main}. We also use a soft-information GRAND variant, Ordered Reliability Bits Guessing Random Additive Noise Decoding (ORBGRAND) \cite{duffy2021ordered}, which is well suited to short, high-rate codes.

In Figure~\ref{fig:sim} we show the BER and BLER vs Eb/N0 for an RLC [128,116], a CA-Polar [128,116] code, and an AES as error-correcting code with $k=116$ and with a zero padding of $n-k = 12$. The rate for all the tested codes presented in this figure is $R=0.91$. The black solid and dashed lines show the performance of RLC codes with GRAND and ORBGRAND, respectively. The red and blue lines in the same present the performance of RLC and CA-Polar codes with CRC, respectively. The yellow lines show the performance of AES as an error-correcting code. All the schemes tested obtain almost the same performance for decoding with a soft-information decoder as proposed by ORBGRAND. Using GRAND for hard-decision decoding, both the RLC code and AES as an error-correcting code achieve the same performance as presented in Section~\ref{sec:cry_ec}. However, we note that the coding schemes with CRC and using GRAND decoder can obtain slightly better performance in SNR higher than $\sim8$dB. This is settled with the known results in the literature, which show that structured codes can slightly outperform the performance of random codes in the tested rate and code-size regime when one utilizes a hard-decision decoder.

\ifLong
\section{Operation Modes for AES as Error Correction}\label{sec:modes}

The US National Institute of Standards and Technology (NIST) provided a list of approved cipher modes of operation for AES \cite{pub2001197}. The simplest of the encryption modes is the electronic codebook (ECB) mode. This is the mode we considered and evaluated in Section~\ref{sec:main}. However, the main drawback of this mode is that it does not hide raw-data patterns, as illustrated in Figure~\ref{fig:ConfNew}.\footnote{We note that the random appearance of the image on the right is indicative, but does not guarantee, that the data has been securely encrypted.} This is because the output of the ECB mode is the same whenever a message is repeated. To avoid this drawback, other cipher modes were proposed by NIST, e.g., Cipher Block Chaining (CBC), Output Feedback (OFB), and Counter (CTR) modes \cite{dworkin2001recommendation}.

\begin{figure*}[ht]				
	\centering	
	\begin{subfigure}[b]{.3\linewidth}
		\includegraphics[trim={1.5cm 2cm 1.5cm 1cm},clip,width=\linewidth, height=1.3in]{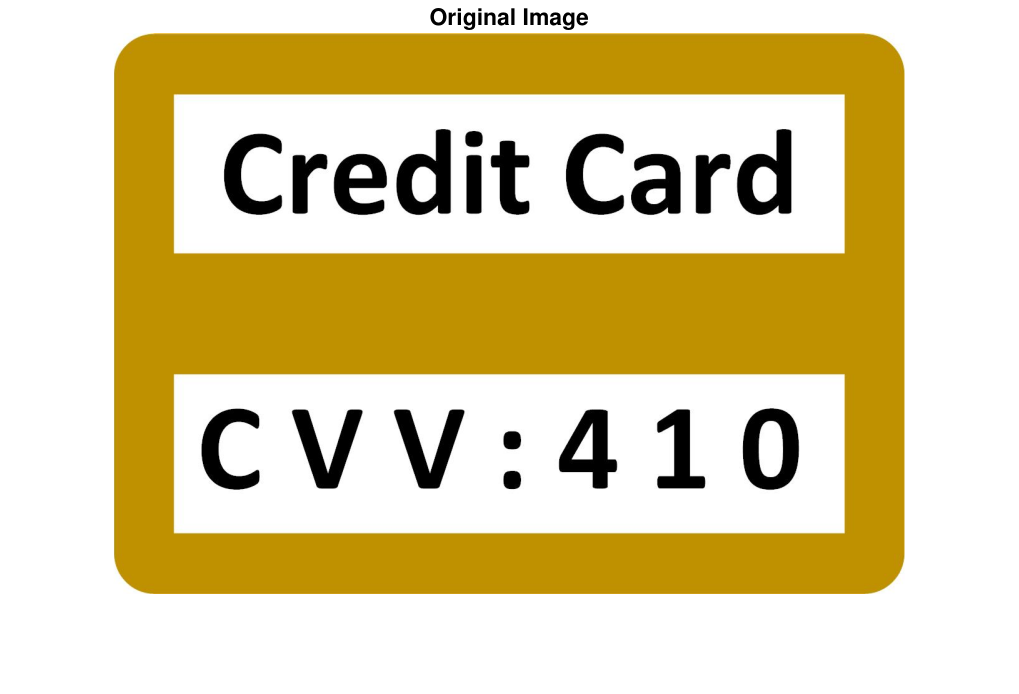}
		\caption{Original image}\label{fig:Conf1}
	\end{subfigure}
	$\quad$
	\begin{subfigure}[b]{.3\linewidth}
		\includegraphics[trim={1.5cm 2cm 1.5cm 1cm},clip,width=\linewidth, height=1.3in]{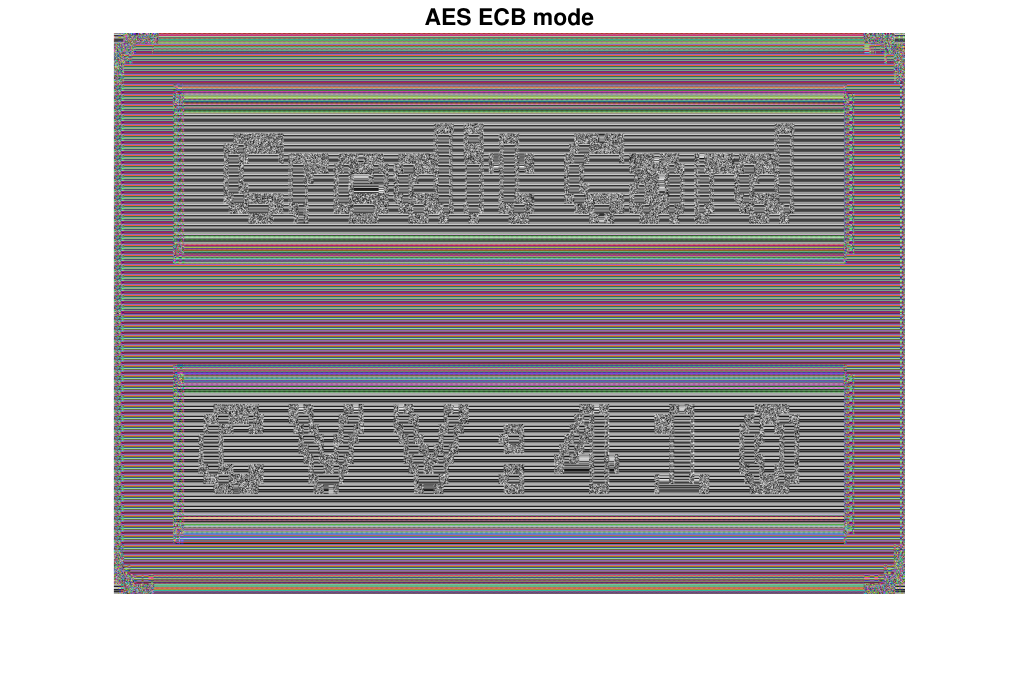}
		\caption{Encrypted using ECB mode}\label{fig:Conf2}
	\end{subfigure}
	$\quad$
	\begin{subfigure}[b]{.3\linewidth}
		\includegraphics[trim={1.5cm 2cm 1.5cm 1cm},clip,width=\linewidth, height=1.3in]{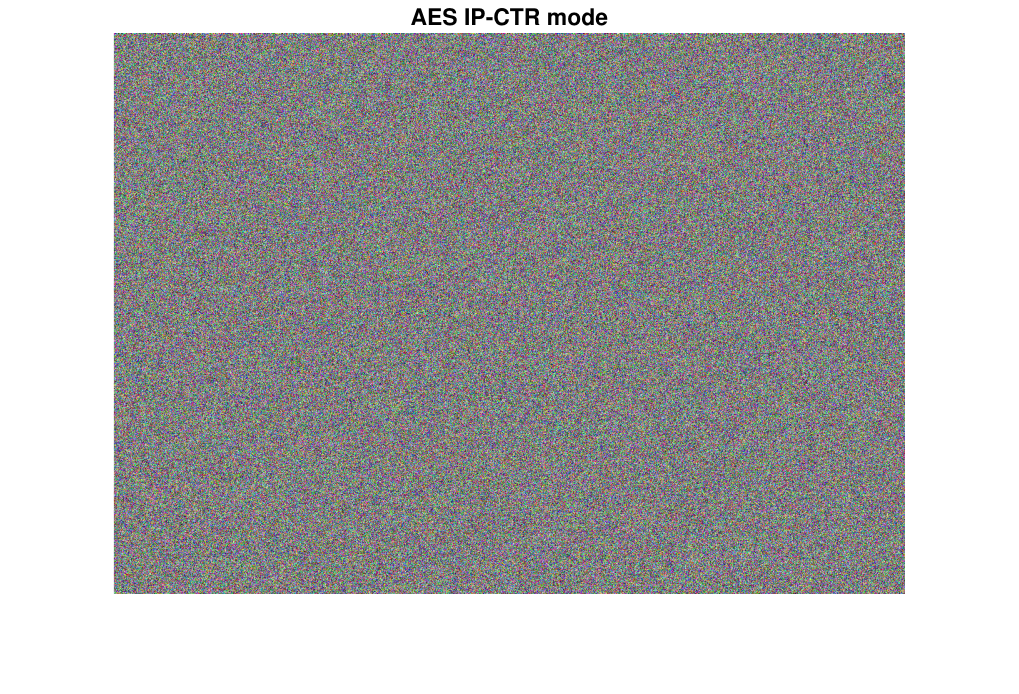}
		\caption{Encrypted using IP-CTR mode}
		\label{fig:Conf3}
	\end{subfigure}	
	\caption{Encryption example using operations modes by AES as a block cipher cryptosystem.}
	\label{fig:ConfNew}
\end{figure*}

Common block cipher modes proposed by NIST suffer from error propagation due to the ciphertext in one iteration, being used as an input in the next. For example, in CBC mode, the ciphertext of the first message being sent is added to the second message before encryption. Thus, any error in decoding the first message is propagated to the second message. This error propagation does not occur for CTR mode. In CTR mode, however, the plaintext itself is not an input of the AES algorithm. Thus, in this case, padding with zeroes does not provide good error correction capabilities.\footnote{One could then rely on using a traditional error correcting code instead. However, one would not then be utilizing the AES algorithm itself to provide the error correction capabilities.}

\begin{figure}
    \centering
    \includegraphics[trim={0cm 0cm 0cm 0cm},clip, width = 1\columnwidth]{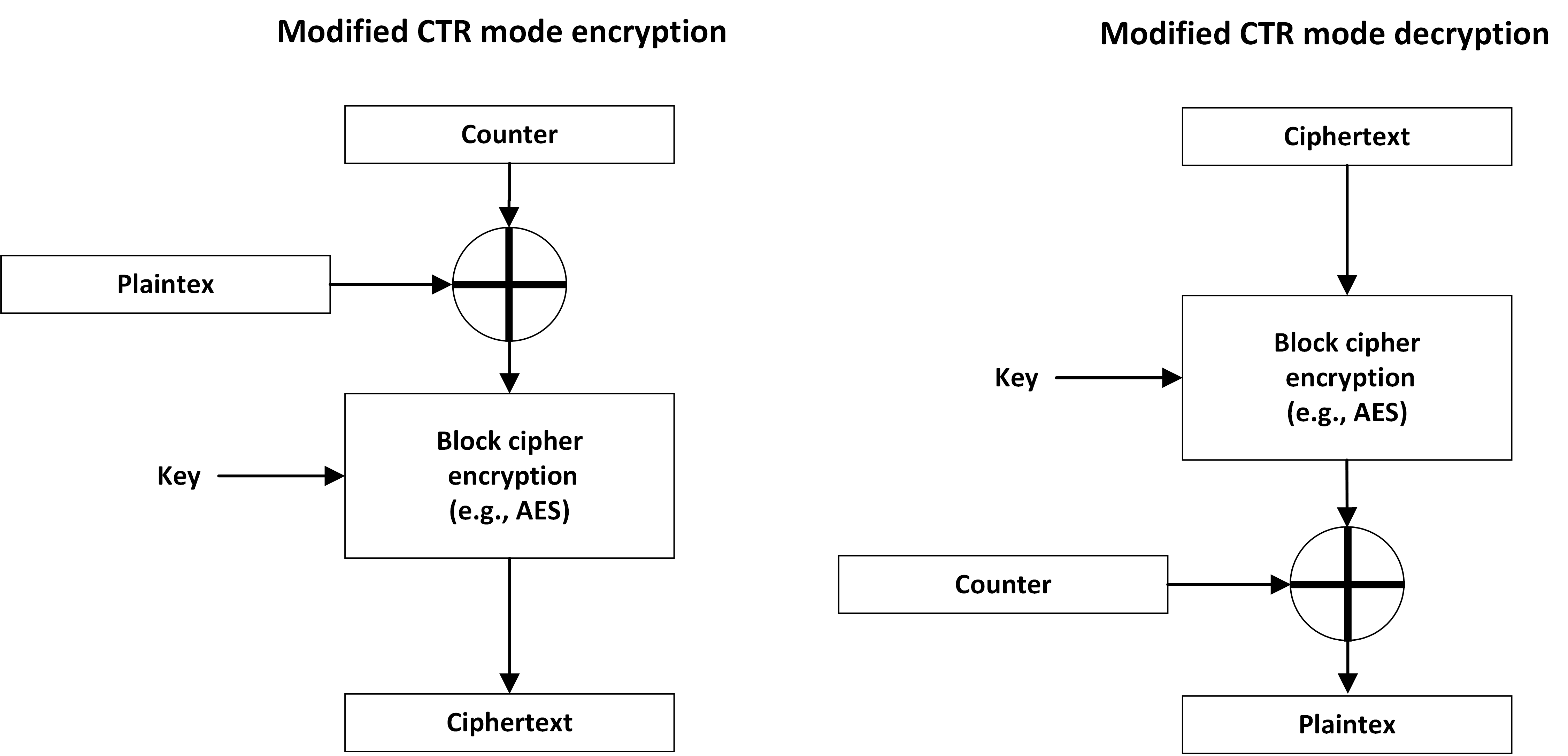}
    \caption{Input Plaintext - Counter (IP-CTR) mode.}
    \label{fig:IPCTR}
\end{figure}

In order to avoid these two issues, error propagation, and the plaintext being used as an input to AES, we present a modified version of
CTR mode. In the original CTR mode, there is a shared counter between the legitimate sender and receiver. The initiated counter number is shared before the first block and is then modified independently by the legitimate sender and receiver before each block transmission. In the encryption process, the counter number is encrypted by the AES, and the outcome is summed with the plaintext. In the decryption, the encrypted counter number is added to ciphertext in order to obtain the original plaintext. The modified version of CTR mode we propose, called input plaintext counter (IP-CTR) mode, works by adding the counter to the plaintext before encrypting them both with the AES algorithm. We illustrate IP-CTR mode in Figure~\ref{fig:IPCTR}. Thus, IP-CTR mode avoids both issues mentioned above. We note that, as shown in Figure~\ref{fig:IPCTR_sim}, in IP-CTR mode we obtain the same error correcting performance as shown in Section~\ref{subsec:sim}.
\begin{figure}
    \centering
    \includegraphics[trim={0cm 0cm 0cm 0cm},clip, width = 1\columnwidth]{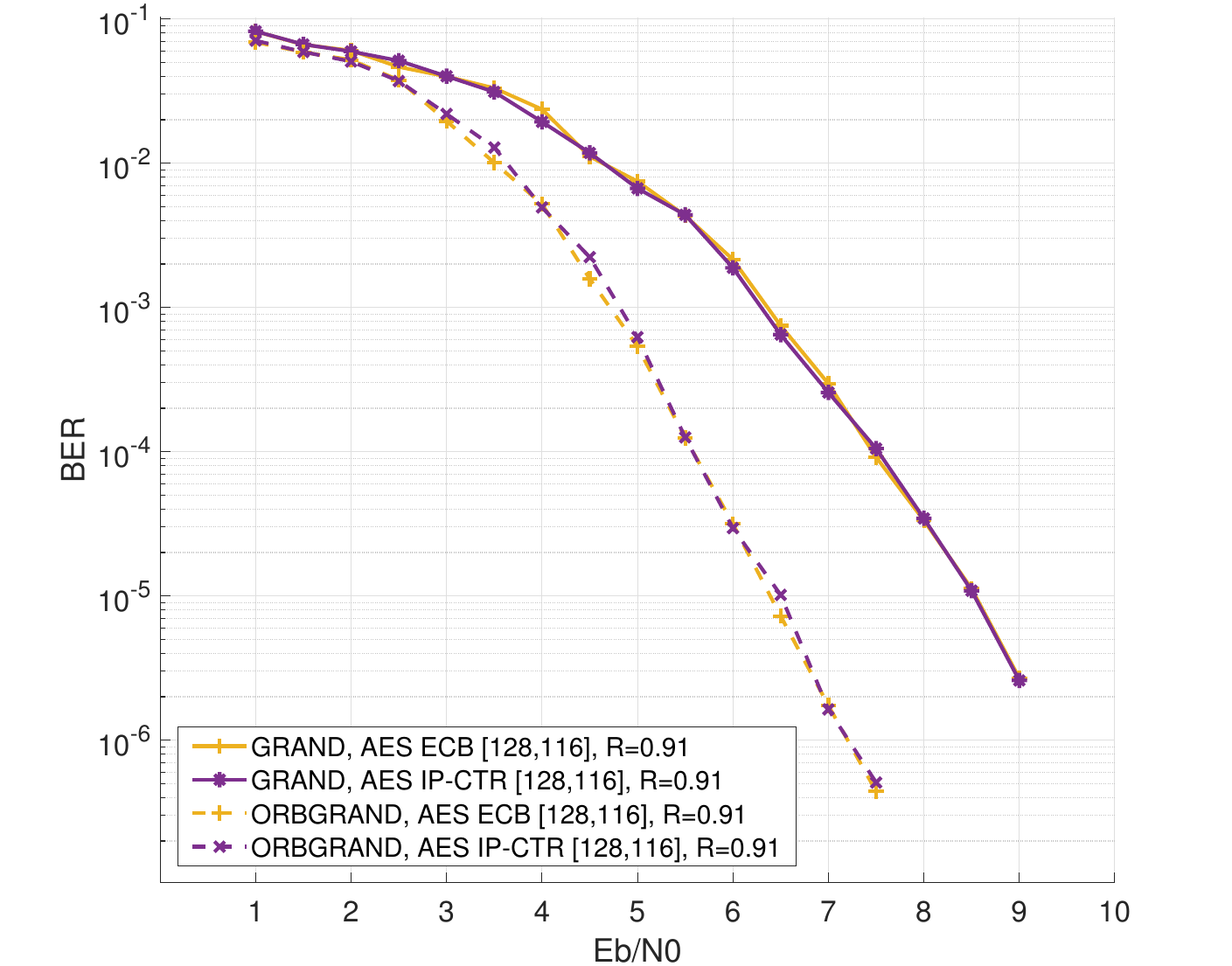}
    \caption{A comparison of AES as an error-correcting code between ECB and IP-CTR operation modes for the same case is presented in Figure~\ref{fig:sim}.}
    \label{fig:IPCTR_sim}
\end{figure}
\else\fi

\section{Future Work}\label{sec:conc}

Our approach opens a new area of exploration. The CEC scheme we present is a good candidate for the investigation of possible security properties. Future work includes considering how to incorporate other cryptosystems in the literature, in order to jointly obtain security and error correction capabilities.
\bibliographystyle{IEEEtran}
\bibliography{ref}

\begin{thebibliography}{10}
\providecommand{\url}[1]{#1}
\csname url@samestyle\endcsname
\providecommand{\newblock}{\relax}
\providecommand{\bibinfo}[2]{#2}
\providecommand{\BIBentrySTDinterwordspacing}{\spaceskip=0pt\relax}
\providecommand{\BIBentryALTinterwordstretchfactor}{4}
\providecommand{\BIBentryALTinterwordspacing}{\spaceskip=\fontdimen2\font plus
\BIBentryALTinterwordstretchfactor\fontdimen3\font minus
  \fontdimen4\font\relax}
\providecommand{\BIBforeignlanguage}[2]{{%
\expandafter\ifx\csname l@#1\endcsname\relax
\typeout{** WARNING: IEEEtran.bst: No hyphenation pattern has been}%
\typeout{** loaded for the language `#1'. Using the pattern for}%
\typeout{** the default language instead.}%
\else
\language=\csname l@#1\endcsname
\fi
#2}}
\providecommand{\BIBdecl}{\relax}
\BIBdecl

\bibitem{shannon1948mathematical}
C.~E. Shannon, ``A mathematical theory of communication,'' \emph{The Bell
  system technical journal}, vol.~27, no.~3, pp. 379--423, 1948.

\bibitem{lin20215g}
X.~Lin and N.~Lee, \emph{5G and Beyond}.\hskip 1em plus 0.5em minus 0.4em\relax
  Springer, 2021.

\bibitem{huffman2021concise}
W.~C. Huffman, J.-L. Kim, and P.~Sol{\'e}, \emph{Concise Encyclopedia of Coding
  Theory}.\hskip 1em plus 0.5em minus 0.4em\relax Chapman and Hall/CRC, 2021.

\bibitem{cover2012elements}
T.~M. Cover and J.~A. Thomas, \emph{Elements of information theory}.\hskip 1em
  plus 0.5em minus 0.4em\relax John Wiley \& Sons, 2012.

\bibitem{dworkin2001advanced}
M.~J. Dworkin, E.~B. Barker, J.~R. Nechvatal, J.~Foti, L.~E. Bassham,
  E.~Roback, J.~F. Dray~Jr \emph{et~al.}, ``Advanced encryption standard
  ({AES}),'' 2001.

\bibitem{nechvatal2001report}
J.~Nechvatal, E.~Barker, L.~Bassham, W.~Burr, M.~Dworkin, J.~Foti, and
  E.~Roback, ``Report on the development of the advanced encryption standard
  ({AES}),'' \emph{Journal of Research of the National Institute of Standards
  and Technology}, vol. 106, no.~3, p. 511, 2001.

\bibitem{hoffstein2008introduction}
J.~Hoffstein, J.~Pipher, J.~H. Silverman, and J.~H. Silverman, \emph{An
  introduction to mathematical cryptography}.\hskip 1em plus 0.5em minus
  0.4em\relax Springer, 2008, vol.~1.

\bibitem{stallings2006cryptography}
W.~Stallings, \emph{Cryptography and network security, 4/E}.\hskip 1em plus
  0.5em minus 0.4em\relax Pearson Education India, 2006.

\bibitem{cohen2022partial}
A.~Cohen, R.~G.~L. D'Oliveira, K.~R. Duffy, and M.~M{\'e}dard, ``Partial
  encryption after encoding for security and reliability in data systems,''
  \emph{arXiv preprint arXiv:2202.03002}, 2022.

\bibitem{moller2004public}
B.~M{\"o}ller, ``A public-key encryption scheme with pseudo-random
  ciphertexts,'' in \emph{European Symposium on Research in Computer
  Security}.\hskip 1em plus 0.5em minus 0.4em\relax Springer, 2004, pp.
  335--351.

\bibitem{duffy2019capacity}
K.~R. Duffy, J.~Li, and M.~M{\'e}dard, ``{Capacity-achieving Guessing Random
  Additive Noise Decoding},'' \emph{IEEE Tran. Inf. Theory}, vol.~65, no.~7,
  pp. 4023--4040, 2019.

\bibitem{duffy19GRAND}
K.~R. {Duffy}, J.~{Li}, and M.~{M\'edard}, ``Capacity-achieving guessing random
  additive noise decoding,'' \emph{IEEE Trans. Inf. Theory}, vol.~65, no.~7,
  pp. 4023--4040, 2019.

\bibitem{duffy2021ordered}
K.~R. Duffy, ``Ordered reliability bits guessing random additive noise
  decoding,'' in \emph{ICASSP 2021-2021 IEEE International Conference on
  Acoustics, Speech and Signal Processing (ICASSP)}.\hskip 1em plus 0.5em minus
  0.4em\relax IEEE, 2021, pp. 8268--8272.

\bibitem{gallager1973random}
R.~Gallager, ``{The random coding bound is tight for the average code
  (corresp.)},'' \emph{IEEE Tran. Inf. Theory}, vol.~19, no.~2, pp. 244--246,
  1973.

\bibitem{tal2015list}
I.~Tal and A.~Vardy, ``List decoding of polar codes,'' \emph{IEEE Transactions
  on Information Theory}, vol.~61, no.~5, pp. 2213--2226, 2015.

\bibitem{ferguson2011cryptography}
N.~Ferguson, B.~Schneier, and T.~Kohno, \emph{Cryptography engineering: design
  principles and practical applications}.\hskip 1em plus 0.5em minus
  0.4em\relax John Wiley \& Sons, 2011.

\bibitem{goldwasser2008lecture}
S.~Goldwasser and M.~Bellare, ``Lecture notes on cryptography,'' 2008.

\bibitem{1056514}
R.~Ahlswede and G.~Dueck, ``Good codes can be produced by a few permutations,''
  \emph{IEEE Transactions on Information Theory}, vol.~28, no.~3, pp. 430--443,
  1982.

\bibitem{tamir2020large}
R.~Tamir, N.~Merhav, N.~Weinberger, and A.~G. i~Fabregas, ``Large deviations
  behavior of the logarithmic error probability of random codes,'' \emph{IEEE
  Transactions on Information Theory}, vol.~66, no.~11, pp. 6635--6659, 2020.

\bibitem{peterson1961cyclic}
W.~W. Peterson and D.~T. Brown, ``Cyclic codes for error detection,''
  \emph{Proceedings of the IRE}, vol.~49, no.~1, pp. 228--235, 1961.

\bibitem{pub2001197}
N.~F. Pub, ``197: Advanced encryption standard ({AES}),'' \emph{Federal
  information processing standards publication}, vol. 197, no. 441, p. 0311,
  2001.

\bibitem{duffy2018guessing}
K.~R. Duffy, J.~Li, and M.~M{\'e}dard, ``{Guessing noise, not code-words},'' in
  \emph{IEEE Int. Symp. Inf. Theory}, 2018, pp. 671--675.

\bibitem{tal2011list}
I.~Tal and A.~Vardy, ``{List decoding of {Polar} codes},'' in \emph{IEEE Int.
  Symp. Inf. Theory}, 2011, pp. 1--5.

\bibitem{arikan2008channel}
E.~Arikan, ``{Channel polarization: A method for constructing
  capacity-achieving codes},'' in \emph{IEEE Int. Symp. Inf. Theory}, 2008, pp.
  1173--1177.

\bibitem{arikan2009channel}
------, ``{Channel polarization: A method for constructing capacity-achieving
  codes for symmetric binary-input memoryless channels},'' \emph{IEEE Tran.
  Inf. Theory}, vol.~55, no.~7, pp. 3051--3073, 2009.

\bibitem{3gppcapolar1}
``{3{GPP} {TS} 38.212, 5{G}-{NR}-{M}ultiplexing and channel coding},''
  \url{https://www.etsi.org/deliver/etsi_ts/138200_138299/138212/15.02.00_60/ts_138212v150200p.pdf},
  [Online].

\bibitem{dworkin2001recommendation}
M.~Dworkin, ``Recommendation for block cipher modes of operation. methods and
  techniques,'' National Inst of Standards and Technology Gaithersburg MD
  Computer security Div, Tech. Rep., 2001.

\end{thebibliography}

\end{document}